# Non-stationary magnetization dynamics driven by spin transfer torque

G. Siracusano[1], G. Finocchio[1,*], A. La Corte[2], G. Consolo[1], L. Torres[3] and B. Azzerboni[1]

[1]Dipartimento di Fisica della Materia e Ingegneria Elettronica, University of Messina, Salita Sperone 31, 98166 Messina, Italy

[2]Dipartimento di Ingegneria Informatica e delle Telecomunicazioni. University of Catania. Viale Andrea Doria, 6 – 95125 Catania, Italy.

[3]Departamento de Fisica Aplicada, University of Salamanca, Plaza de la Merced s/n, 37008 Salamanca, Spain

**Abstract-** This paper shows that the presence of two dynamical regimes, characterized by different precessional-axis, is the origin of the non-monotonic behavior of the output integrated power for large-amplitude magnetization precession driven by spin-polarized current in nanoscale exchange biased spin-valves. In particular, at the transition current between those two regimes exists an abruptly loss in the integrated output power. After the introduction of a time-frequency analysis of magnetization dynamics based on the wavelet transform, we performed a numerical experiment by means of micromagnetic simulations. Our results predicted that, together with a modulation of the frequency of the main excited mode of the magnetization precession, at high non-linear dynamical regime the instantaneous output power of the spin-torque oscillator can disappear and then reappear at nanosecond scale.

# I. INTRODUCTION

The discovery that a spin-polarized current interacting with a nanomagnet can produce several different types of magnetic dynamics[1,2] open perspectives in applied physics for spintronic technology.[3,4,5,6,7,8] These offers the possibility of applications that include at least magnetoresistive random access memories,[4] nano-oscillators,[7,9] and radiofrequency detectors.[10] Frequency[7] and time[8] domain measurements of magneto-resistance signal in the "state of the art" spintronic devices (spin valves,[11] magnetic tunnel junctions,[12] point contact geometries[13]) show very rich dynamical stability diagrams with switching between static magnetic states and different steady-state precession characterized by uniform and non-uniform magnetization pattern. In particular, the frequency, the linewidth, and the microwave output power of the precessions show strongly dependence on external field and current.[14]

In addition, exchange bias nanoscale spin-valves with a Py-free layer (Py= $Ni_{80}Fe_{20}$) of elliptical cross-sectional area exhibits dynamics with series of jumps in frequency between stationary nonlinear modes characterized by either different spatial distribution[15,16] and different Hausdorff dimension.[17] Those measurements also show, for some values of current and field, a non-stationary magnetization dynamics related to nanosecond switching between a dynamical mode and a static magnetic configuration or between different dynamical modes.[16] In the latter case, this non-stationary regime is characterized by a spectrum with two well defined peaks in frequency, and it is observed before that large-amplitude magnetic precession is driven, or when a device is biased near the boundary between the jumps of two different modes. In the large amplitude dynamical regime, while the frequency of the main excited mode monotonically decreases as function of the current, the integrated output power shows non-monotonic behaviour with a well defined minimum at least for the device studied in Ref[15].

Here, we first performed a numerical experiment based on the solution of the Landau-Lifshitz-Gilbert-Slonczewski[18,19] (LLGS) equation in order to identify the origin of the minimum in the integrated output power. Secondly, we introduce a continuous wavelet analysis[20] of non-stationary magnetization dynamics driven by spin-polarized current, and reproduce the results of non-stationary regime of experimental direct time-domain electrical measurements in nanoscale exchange biased spin-valves obtained for a four-nanosecond windowed Fourier transform. Finally, we predicted by combining micromagnetic simulations and wavelet analysis, that the excited modes of a spin-torque nano-oscillator show together to a frequency modulation[21] a nanosecond intermittent disappearing and reappearing of the instantaneous microwave output power.

## II. NUMERICAL DETAILS

We simulate exchange biased spin valves Py(4nm)(free layer)/Cu(8nm)/Py(4nm)(pinned layer)/Ir$_{20}$Mn$_{80}$/ (8nm) with elliptical cross sectional area (130 nm x 60nm) in the same experimental framework of Ref[16]. We use a saturation magnetization $M_S$=650x10$^3$ A/m, a free-layer damping $\alpha_F$=0.025, a pinned layer damping $\alpha_P$=0.2, an exchange constant of $A$=1.3 10$^{-11}$ J/m.[22] For the spin-torque efficiency $\varepsilon(\theta)$ and the magneto-resistance $r(\theta)$, we use the formulation developed by Slonczewski[23] for symmetric spin-valves

$$\varepsilon(\theta) = 0.5 P \Lambda^2 / \left(1 + \Lambda^2 + (1-\Lambda^2)\cos(\theta)\right) \quad \text{and} \quad r(\theta) = \left(1 - \cos^2(\theta/2)\right) / \left(1 + \chi \cos^2(\theta/2)\right), \quad \text{where}$$

$\Lambda^2 = \chi + 1$, $\chi$ is the giant-magneto-resistance asymmetry parameter, $P$ is the current spin-polarization factor, the parameters values $\chi = 1.5$ and $P = 0.38$ have been obtained by fitting to the experimentally-measured ensemble-average switching time.[19] The pinned layer is exchange biased in the plane of the sample at an angle of 45° with respect to the major axis of the ellipse, with an effective exchange field of 75 mT. We simulate the entire spin-valve including the effects of the back action of the torque to the pinned layer and a spin torque with a stochastic component (we include thermal fluctuations also in the pinned layer).[22]

## III. RESULTS

In order to qualitatively explain the non-monotonic behaviour of the integrated output power and to compare our numerical results directly to the experimental data (Ref. [15], Fig. 7), we study large amplitude magnetization dynamics for an external field $H_{ext}$=68mT applied along -45° with respect to the easy axis of the ellipse. We identify the proportionality factor ($\kappa$)[15] between experimental and applied current ($I$) by fitting the frequency versus current data, we computed $\kappa = 0.34$. The integrated power signal amplitude shows a minimum for a current around 7.5mA which does not correspond to any non-stationary region or jump between different non-linear dynamical modes. Near this current value, our results show two different dynamical regimes, those are characterized by a magnetization oscillation axis (**m**$_f$ in Fig. 1(a)) at an angle between 0 and -45° (regime A, lower current) or between -45° and -125° (regime B, larger current). Figure 1 shows the time evolution of the normalized average magnetization (<$m_X$> (a) and <$m_Y$> (b) (c)) for $\kappa I = 6.3$mA in regime A (frequency of the main mode $f$=3.3GHz) and $\kappa I = 8$mA in regime B ($f$=2.8GHz). These magnetization time-traces are qualitatively similar to the real-time magnetoresistance signal measured via a microwave storage oscilloscope and displayed in Figs. 4(b) and (c) of Ref[16]. For large-amplitude magnetization dynamics, first the integrated output power increases as function of current (regime A) then it decreases abruptly (in the transition from regime A to regime B), finally it increases as function of current (regime B). Simulations performed considering the pinned layer fixed did not show these two dynamical regimes, we argue that this is the main point which gives the different results between our micromagnetic simulations and the ones published in Ref[15].

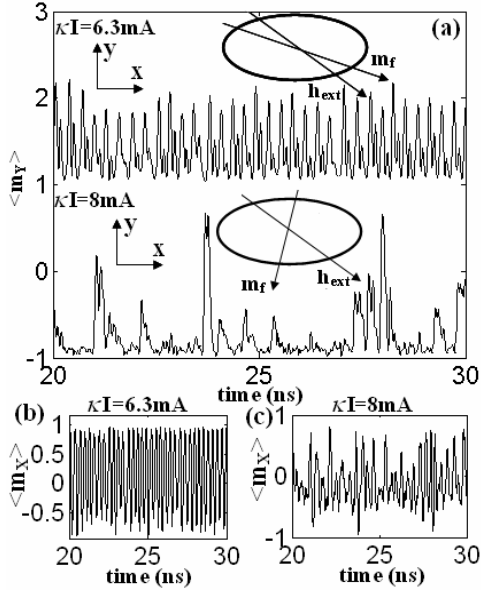

FIG. 1(a) Temporal evolution of normalized y-component of the average magnetization for $\kappa I$=6.3mA (top, an offset of two is applied) and $\kappa I$=8mA (bottom). The axis of the magnetization oscillation ($\mathbf{m}_f$ in figure) is in the former case between 0° and -45° and in the latter case between -45° and -125°. Temporal evolution of normalized x-component of the average magnetization for (a) $\kappa I$=6.3mA and (b) $\kappa I$=8mA.

Concerning the same kind of devices, experimental data published in Ref. [16] show that non-stationary magnetization dynamics is driven before of the large-amplitude magnetization precession. In particular, for $I$=4.5mA and $H_{ext}$=60mT, the power spectrum of the real-time voltage signal (for a signal of 20ns see Fig. 6(e) in Ref[16]) captured via microwave storage oscilloscope (the voltage is directly proportional to the magneto-resistive signal) shows two excited modes $P_1$ and $P_2$ ($f_{P1}$=3.9GHz and $f_{P2}$=4.6GHz). By performing the Fourier transform with a window time of 4ns it can be observed either $P_1$ or $P_2$ (see Figs. 6(f) and 6(g) in Ref[16]), this result shows the non-periodic origin of this magnetization precession; furthermore the presence of those two modes at second scale also shows their non-transient origin (see Fig. 6(d) in Ref. [16]). From computational point of view, it is important to find out a tool which systematically gives information about the time localization of the excited modes. We use a wavelet-based analysis (the wavelet is the natural generalization of the Windowed Fourier transform) and differently by other approaches, we systematically identify the scale set directly from the power spectrum related to *r(t)* (magnetoresistance time-domain signal). The continuous wavelet transform of a function *r*(t) (we consider the time-dependence of the magneto resistance signal) is a linear transform $W_r(u,s)$ given by[20]

$$W_r(u,s) = \frac{1}{\sqrt{s}} \int_{-\infty}^{+\infty} r(t) \psi^* \left( \frac{t-u}{s} \right) dt \qquad (1)$$

being *s* and *u* the scale and translation parameters of the mother wavelet $\psi(t)$, which defines the wavelet family function as $\psi_{u,s}(t) = \frac{1}{\sqrt{s}} \psi \left( \frac{t-u}{s} \right)$. In our study, in order to characterize both amplitude and phase of time domain magneto-resistive signal (*r(t)*) we use the complex Morlet wavelet family $\psi_{u,s} = \frac{1}{\sqrt{s\pi f_B}} e^{j2\pi f_c \left( \frac{t-u}{s} \right)} e^{-\left( \frac{t-u}{s} \right)^2 / f_B}$ with Fourier spectrum $\Psi_{u,s}(f) = \sqrt{s} e^{-\pi^2 f_B (sf - f_c)^2} e^{-j2\pi uf}$, where $f_C$ and $f_B$ are two parameters ($\Psi(f)$ is the Fourier transform of $\psi(t)$). The use of a wavelet analysis allows to characterize a signal in the time-frequency space to study the non-stationary behavior (for a complete review of wavelet theory see Refs. 24, 25). The $f_B$ (called bandwidth parameter) can rule the band of the complex Morlet Fourier spectrum giving narrowed band as it increases, consequently this parameter is correlated to the frequencies we have to analyze independently of each other, furthermore from practical reasons $f_B$ and $f_C$ have to be large enough to made the mean of $\psi(t)$ arbitrarily small.[26]

Our results suggest that the continue wavelet transform $W_r(u,s)$ shows better statistical performance than any other time-frequency analysis methods used to analyze of the simulated signals. Figure 2(a) (monotonic line) shows the normalized integrated power spectrum of the voltage signal (inset of Fig. 2(a)), the slope of this curve increases rapidly close to the high power frequencies $f_{P1}$ and $f_{P2}$ while it increases slowly elsewhere (depending on the noise in the power spectrum). Given a fixed dimension *N* of a scale set $\{s_i\}_{i=1...N}$ for the wavelet family, the *y*-axis is divided in *N*+2 points (Fig. 2(a) shows a decomposition for *N*=9, see the number in the right) and for each point can be obtained a frequency $f_i$ and a scale $s_i$ (e.g. for the number 3, $f_3$=4.44GHz and $s_3 = f_S/f_3$ being $f_S$=30 GHz the sampling frequency).

The best way to show the time-frequency characterization of a signal is the wavelet scalogram (WS), furthermore the integral of the WS of a signal over the time can be correlated to the Fourier spectrum of that signal directly for a fixed scale parameter.[27] The WS of *r(t)* is defined as $P_W^r(t,f) = |W_r(u,s)|^2$, being the time *t* the center value of the wavelet translated by *u*, and the frequency *f* is computed directly by the scale factor as $f = f_S/s$ being $f_S$ the sampling frequency (rigorously speaking $f$ is the scale frequency and in general it is different from the Fourier frequency, but for the complex Morlet wavelet with $f_C=1$ the two frequencies are nearly identical).[28]

Fig. 2(b) shows the WS (arb. units) computed for the voltage signal of the inset of Fig. 2(a), we use the following parameters $N=22$, $f_B=300$, and $f_C = 1$, as can be noted, the results of our computations are consistent with the data displayed in Fig. 6 of Ref. [16].

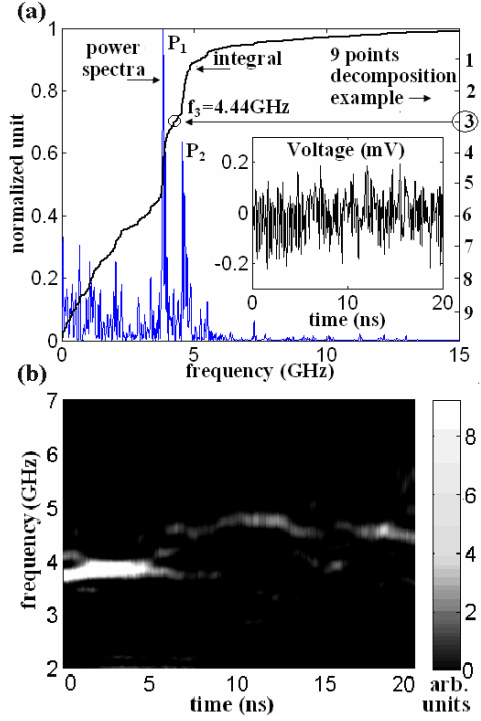

FIGURE 2(a): Normalized power spectrum and integrated power spectrum of voltage signal (inset) measured for $I=4.5$mA and $H=60$mT in Ref[16]. The numbers in the right represent an example of 9 points decomposition of the y-axis to identify the scale set $\{s_i\}_{i=1...9}$ for the wavelet transform. (b) Wavelet scalogram of the voltage signal displayed in the inset of Fig. 2(a).

To perform time-frequency analysis of micromagnetic simulations, we introduce a generalization of the micromagnetic spectral mapping technique (MSMT)[29,30] the micromagnetic WS (MWS). The MWS is the sum over all the computational cells ($N_C$) of the WS of the magneto-resistance temporal evolution computed for each cell $W_r^i(u,s)$.

$$P_W^r(t,f) = \frac{1}{N_C}\sum_{i=1}^{N_C} P_W^{r_i}(t,f) \qquad (2)$$

The scale set $\{s_i\}_{i=1...N_c}$ is determined directly by the spectrum computed with the MSMT with the procedure described above (see Fig.2(a)). In general, this study can be also performed for each component of the magnetization. Figure 3 shows (a) the MWS ($N$=22, $f_B$=10$^3$), (b) the power spectrum computed by the MSMT, and (c) the temporal evolution of the normalized magneto-resistance signal $r(t)$ for the following value of current density ($J$), external field ($H$), and temperature ($T$): $J$=1.5 10$^8$ A/cm$^2$, $H$=60mT applied along -45° with respect to the easy axis, and $T$=300K.

The MWS is able to detect the intermittent features of the $r(t)$ (for example Fig. 3(a) t ≈ 9ns and t ≈ 25ns), a characteristic which can not be find out via Fourier analysis. Our numerical experiment shows several interesting results. The excited modes of a spin-torque nano-oscillator are turned off (for nanosecond interval) in an intermediate static magnetic configuration characterized by output power very small (intermittent non-periodic behaviour) and than re-excited. The magnetization dynamics is characterized of a main mode with a frequency which moves in time in a range of a few hundreds MHz at least, this mechanism sets the minimum value of the linewidth of the mode. This frequency modulation has been also predicted by stochastic non-linear theory.[21] As the current increases, first a region in which the intermittences disappear (see Fig. 3(d)-top which shows the MWS for $J$=2 10$^8$ A/cm$^2$) then a region where the intermittences reappear (see Fig.3(d)-bottom which shows the MWS for $J$=3 10$^8$ A/cm$^2$). This gives rice at least to a linewidth broadening with a Lorentzian shape added to the ideal power spectrum.[31, 32]

This wavelet-based analysis together with micromagnetic simulations is able to investigate completely the time-frequency behaviour of magnetization dynamics from LLGS equation including the transient dynamics in magnetization switching processes and the existence of persistent but non-periodic current-driven magnetic states. From a more general point of view, it can be used to analyze all the physical problems where a systematic non-stationary analysis has to be performed.

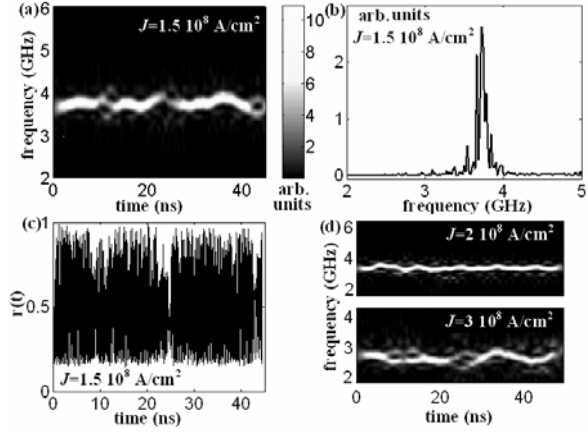

FIGURE 3: Theoretical computations for $J=1.5\ 10^8$ A/cm$^2$, $H$=60mT applied along -45° with respect to the easy axis, and $T$=300K of (a) micromagnetic wavelet scalogram ($N$=22), (b) the power spectrum computed by the micromagnetic spectral mapping technique, and (c) the temporal evolution of the normalized magneto-resistance signal $r(t)$ computed by numerically solving the Landau-Lifshitz-Gilbert-Slonczewski as described in the text. (d) micromagnetic wavelet scalogram computed for same field and temperature of (a) and for (top) $J=2\ 10^8$ A/cm$^2$ and (bottom) $J=3\ 10^8$ A/cm$^2$. The colorbar of (d) is the same of (a).

## IV. CONCLUSION

In conclusion, experiments and simulation of nanoscale exchange biased spin-valves show very rich dynamical behaviour. At high non-linear dynamical regime, even throw the frequency of the main excited mode decreases monotonically as function of current, the integrated output power shows non-monotonic behaviour with a minimum related to a change in the oscillation axis of the magnetization. This is a crucial point to take into account to design spin-torque nano-oscillators.

By combining micromagnetic simulations and time-frequency characterization of the magnetization dynamics, we observe together to a frequency modulation, current-dependent nanosecond intermittent disappearing and reappearing of the magnetization dynamics and the instantaneous microwave output power. This aspect has to taken into account to improve theoretical prediction of linewidth of a spin torque nano-oscillator.

## ACKNOWLEDGMENTS

This work was partially supported by Spanish projects MAT2008-04706/NAN and SA025A08.